\documentstyle[12pt,procsla]{article}
  
  \catcode`\@=11
  \long\def\@makefntext#1{
  \protect\noindent \hbox to 3.2pt {\hskip-.9pt  
  $^{{\ninerm\@thefnmark}}$\hfil}#1\hfill}		%CAN BE USED 
  
  \def\@makefnmark{\hbox to 0pt{$^{\@thefnmark}$\hss}}  %ORIGINAL 
  	
  \def\ps@myheadings{\let\@mkboth\@gobbletwo
  \def\@oddhead{\hbox{}
  \rightmark\hfil\ninerm\thepage}   
  \def\@oddfoot{}\def\@evenhead{\ninerm\thepage\hfil
  \leftmark\hbox{}}\def\@evenfoot{}
  \def\sectionmark##1{}\def\subsectionmark##1{}}
  
\begin{document}

%\preprint{\vbox{baselineskip=14pt
\begin{flushright}
%\rightline{
UH-511-932-99  \\
%} \break
%\rightline{
May 1999  %} }}
\end{flushright}
  
  \centerline{\normalsize\bf EXOTIC EXPLANATIONS FOR NEUTRINO
ANOMALIES\footnote
{Invited talk at the ``8th International Symposium
  on Neutrino Telescopes'', Venice, Feb. 1999.}}  
  \baselineskip=16pt
  %\centerline{\normalsize\bf MANUSCRIPT BY COMPUTER}
  %\centerline{\footnotesize\sf (For subsequent $\sim$ 15\% photoreduction)
  
  %\vfill
  %\vspace*{0.6cm}
  \centerline{\footnotesize SANDIP PAKVASA}
  \baselineskip=13pt
  \centerline{\footnotesize\it University of Hawaii}
  \centerline{\footnotesize\it Department of Physics and Astronomy}
  \baselineskip=12pt
  \centerline{\footnotesize\it Honolulu, HI  96822 USA}
  \centerline{\footnotesize E-mail: pakvasa@uhheph.phys.hawaii.edu}
%  \vspace*{0.3cm}
%  \centerline{\footnotesize and}
%  \vspace*{0.3cm}
%  \centerline{\footnotesize SECOND AUTHOR'S NAME}
%  \baselineskip=13pt
%  \centerline{\footnotesize\it Group, Company, Address, City, State ZIP/Zone,
%  Country}
  
  %\vfill
  \vspace*{0.9cm}
  \abstracts{I review explanations for the three neutrino anomalies
  (solar, atmospheric and LSND) which go beyond the ``conventional''
  neutrino oscillations induced by mass-mixing.  Several of these
  require non-zero neutrino masses as well.}
  %\vspace*{0.6cm}
  \normalsize\baselineskip=15pt
  \setcounter{footnote}{0}
  \renewcommand{\thefootnote}{\alph{footnote}}
  \section{Introduction}
  According to Webster's New English Dictionary\cite{webster}, amongst the meanings of
  the word 
  ``exotic'' are: ``foreign'', ``strangely beautiful'', ``enticing'', ``having the
  charm or fascination of the unfamiliar''. At one time neutrino
  oscillations would have been considered exotic, but today they seem
  mundane and ordinary!  The explanations to be discussed here will be
  definitely ``strange'' and ``unfamiliar'' but may not seem to be 
``charming'' or ``enticing''!

As is well-known, it is not possible to account for all three neutrino
anomalies with just the three known neutrinos $(\nu_e, \nu_\mu$ and
$\nu_\tau)$. If one
of them can be explained in some other way, then no extra sterile neutrinos
need be invoked.  This is one motivation for exotic scenarios.  
It is also important to rule out all explanations other
than oscillations in order to establish neutrino mixing and
oscillations as the unique explanation for the three observed neutrino
anomalies.

I should mention that, in general, some (but not all) non-oscillatory
explanations of the neutrino anomalies will involve non-zero neutrino
masses and mixings.  Therefore, they tend to be neither elegant nor
economical.  But the main issue here is whether we can establish
oscillations unequivocally and uniquely as the cause for the observed
anomalies.

I first summarize some of the exotic scenarios and then 
consider each anomaly in turn.

\section{Mixing and Oscillations of massless $\nu's$}

There are three different ways that massless neutrinos may mix and even
oscillate.  These are as follows.

1. If flavor states are mixtures of massless as well massive states, then
when they are produced in reactions with Q-values smaller than the
massive state; the flavor states are massless but not
orthogonal\cite{lee}.  
For example, if
$\quad  \nu_e = \sum_{i=1}^{4}U_{ei}\nu_i$, 
\noindent
$\nu_\mu=\sum_{i=1}^{4} U_{\mu i}
\nu_i$ and $m_i = 0$ for $i= 1$ to 3 but $m_4$ = 50 GeV; then ``$\nu_e$''
produced in $\beta$-decay and ``$\nu_\mu$'' produced in $\pi$-decay
are massless but not orthogonal and
\begin{equation}
\left < \nu_e \mid \nu_\mu \right >  =  - U_{e4}^* \ U_{\mu 4}
\end{equation}

On the other hand $\nu_e$ and $\nu_\mu$ produced in $W$ decay will be
not massless and will be more nearly orthogonal.  Hence, the definition of
flavor eigenstate is energy and reaction dependent and not fundamental.
Current limits\cite{langacker} on orthogonality of $\nu_e, \nu_\mu$ and $\nu_\tau$ make
it impossible for this to play any role in the current neutrino
anomalies.

2.  When Flavor Changing Neutral Currents as well as Non-Universal
    Neutral Current Couplings of neutrinos exist:
\begin{eqnarray}
& &\epsilon_q \frac{4 G_F}{\sqrt{2}} \left \{ 
\bar{\nu}_{el} \ \gamma_\mu \ \nu{_{\tau_{L}}} \bar{q}_L \ \gamma_\mu q_L \ \ +
h.c. \right \} \\ \nonumber
& + & \epsilon'_q \frac{4 G_F}{\sqrt{2}} \ \left \{ 
\bar{\nu}_{e  L} \ \gamma_\mu \ \nu{_{e_{L}}} - \bar{\nu}_{\tau L} 
\gamma_{\mu} \ \nu_{\tau L} \right \}  \left \{ \bar{q}_L \ \gamma_\mu \ q_L
\right \}
\end{eqnarray}
Such couplings arise in R-parity violating supersymmetric theories\cite{roulet}.
In this case the propagation of $\nu_e$ and $\nu_\tau$ in matter is
described by the equation:
\begin{eqnarray}
i \frac{d}{dt}
\left (
\begin{array}{c}
\nu_e \\
\nu_\tau
\end{array}
\right )_L 
=
\left (
\begin{array}{cc}
\alpha + \beta & \gamma \\
\gamma         & -\beta
\end{array}
\right )
\left (
\begin{array}{c}
\nu_e \\
\nu_\tau
\end{array}
\right )_L
\end{eqnarray}
where $\alpha =G_F N_e, \beta = \epsilon_q G_F N_q$ and
$\gamma = \epsilon'_q G_F N_q$.  There is a resonance at
$\alpha + 2 \beta = 0$    or $\epsilon_q = -\frac{1}{2}
N_e/N_q$
and $\nu_e$ can convert to $\nu_\tau$ completely.  This matter effect
which effectively mixes flavors in absence of masses was first pointed
out in the same Wolfenstein paper\cite{wolfenstein} where matter effects were discussed
in general.  The resonant conversion is just like in the conventional
MSW effect, except that there is no energy dependence and $\nu$ and $\bar{\nu}$
are affected the same way.  This possibility has been discussed in connection with both
solar and atmospheric anomalies.

3.  (a) Flavor violating Gravity wherein it is proposed that
    gravitational couplings of neutrinos are flavor
    non-diagonal\cite{gasperini} and equivalence principle is violated.  For
    example, $\nu_1$ and $\nu_2$ may couple to gravity with different
    strengths:

\begin{equation}
H_{gr} = f_1 GE \phi + f_2 GE \phi
\end{equation}
where $\phi$ is the gravitational potential.  Then if $\nu_e$ and
$\nu_\mu$ are mixtures of $\nu_1$ and $\nu_2$ with a mixing angle
$\theta$, oscillations will occur with a flavor survival probability
\begin{equation}
P= 1- sin^2 2 \theta sin^2 ( \frac{1}{2} \delta f \phi EL)
\end{equation}
when $\phi$ is constant over the distance L and $\delta f=f_1-f_2$ is
the small deviation from universality of gravitational coupling.
Equivalence principle is also violated.

(b) Another possibility is violation of Lorentz invariance\cite{coleman} wherein all
particles have their own maximum attainable velocities (MAV) which are all
different and in general also different from speed of light:  Then if
$\nu_1$ and $\nu_2$ are MAV eigenstates with MAV's $v_1$ and
$v_2$ and $\nu_e$ and $\nu_\mu$ are mixtures of $\nu_1$ and $\nu_2$
with mixing angle $\theta$,
the survival probability of a given flavor is
\begin{equation}
P=1-sin^2 2 \theta \ sin^2 \ \left (\frac{1}{2} \delta v EL \right )
\end{equation}
where $\delta v = v_1-v_2$.

As far as neutrino oscillations are concerned, these two cases are
identical in their dependence on LE instead of L/E as in the
conventional oscillations\cite{glashow}.

\section{Neutrino Decay\cite{discussion}:}

Neutrino decay implies a non-zero mass difference between two neutrino
states and thus, in general, mixing as well.  We assume a
component of $\nu_\alpha,$ i.e., $\nu_2$, to be the only unstable state,
with a rest-frame lifetime $\tau_0$, and we assume two flavor mixing,
for simplicity:
\begin{equation}
\nu_\mu = cos \theta \nu_2 \ + sin \theta \nu_1
\end{equation}
with $m_2 > m_1$.  From Eq. (2) with an unstable $\nu_2$, the $\nu_\alpha$
survival probability is
\begin{eqnarray}
P_{\alpha \alpha} &=& sin^4 \theta \ + cos^4 \theta {\rm exp} (-\alpha L/E)
            \\ \nonumber
&+& 2 sin^2 \theta cos^2 \theta {\rm exp} (-\alpha L/2E)
            cos (\delta m^2 L/2E),
\end{eqnarray}
where $\delta m^2 = m^2_2 - m_1^2$ and $\alpha = m_2/ \tau_0$.
Since we are attempting to explain neutrino data without oscillation
there are two appropriate limits of interest.  One is when the $\delta
m^2$ is so large that the cosine term averages to 0.  Then the survival
probability becomes
\begin{equation}
P_{\mu\mu} = sin^4 \theta \ + cos^4 \theta {\rm exp} (-\alpha L/E)
\end{equation}
Let this be called decay scenario A.  The other possibility is when
$\delta m^2$ is so small that the cosine term is 1, leading to a
survival probability of 
\begin{equation}
P_{\mu \mu} = (sin^2 \theta + cos^2 \theta {\rm exp} (-\alpha L/2E))^2
\end{equation}
corresponding to decay scenario B.  Decay models for both kinds of
scenarios can be constructed; although they require fine tuning and are
not particularly elegant.

\section{LSND}

In the LSND experiment, what is observed is the following\cite{athana}.  In the decay
at rest (DAR) which is $\mu^+ \rightarrow e^+ \nu_e \bar{\nu}_\mu$ 
which should give a pure
$\nu_e$ signal, there is a flux of $\bar{\nu_e's}$ at a level of about
$3.10^{-3}$ of the $\nu_e's$.  (There is a similar
signal for $\nu'_es$ accompanying $\bar{\nu}'_es$ in the decay in flight
of $\mu^- \rightarrow e^- \bar{\nu}_e \nu_\mu)$.  Now this could be
accounted for without oscillations\cite{some} provided that the conventional decay mode
$\mu^+ \rightarrow e^+ \nu_e \bar{\nu}_\mu (\mu^- \rightarrow
e^- \bar{\nu}_e \nu_\mu)$ is accompanied by the rare mode 
$\mu^+ \rightarrow e^+ \bar{\nu}_e {\rm X} \ ( \mu^- \rightarrow e^- \nu_e
\bar{\rm X})$ at a level of branching fraction of $3.10^{-3}$.  Assuming X
to be a single particle, what can X be?  It is straight forward to
rule out X as being (i) $\nu_\mu$ (too large a rate for
Muonium-Antimuonium transition rate), (ii) $\nu_e$ (too large a rate
for FCNC decays of
$Z$ such as $Z \rightarrow \mu\bar{e} + \bar{\mu} e)$ and (iii)
$\nu_\tau$ (too large a rate for FCNC decays of $\tau$ such as $\tau
\rightarrow \mu ee)$.

The remaining possibilities for X are $\bar{\nu}_\alpha$ or
$\nu_{\rm{sterile}}$.
No simple models exist which lead to such decays.  Rather Baroque models
can be constructed which involve a large number of new particles\cite{grossman}.

Experimental tests to distinguish this rare decay possibility from the
conventional oscillation explanation are easy to state.  In the rare decay
case, the rate is constant and shows no dependence on L or E.

	One can also ask whether the $\bar{\nu}_e$ events seen in LSND
could have been caused by new physics at the detector; for example a
small rate for the ``forbidden'' reaction $\bar{\nu}_\mu + p \rightarrow
n + e^+$.  This is ruled out, since a fractional rate of $3.10^{-3}$ for
this reaction, leads (via crossing) to a rate for the decay mode $\pi^+
\rightarrow e^+ \nu_\mu$ in excess of known bounds.

\section{Solar Neutrinos}

Oscillations of massless neutrinos via Flavor Changing Neutral Currents
(FCNC) and Non Universal Neutral Currents (NUNC) in matter have
been considered\cite{roulet,bahcall} as explanation for the solar neutrino observations, most
recently by Babu, Grossman and Krastev\cite{krastev}.  Using the most recent data from
Homestake, SAGE, GALLEX and Super-Kamiokande, they find good fits with
$\epsilon_u \sim 10^{-2}$ and $\epsilon'_u \sim 0.43$ or $\epsilon_d
\sim$ 0.1 to 0.01 with $\epsilon'_d \sim 0.57$.

Since the matter effect in this case is energy independent, the
explanation for different suppression for $^8B$, $^7Be$ and pp neutrinos
as inferred is interesting.  The differing suppression arises from the
fact the production region for each of these neutrinos differ in
electron and nuclear densities.  This solution resembles the large angle
MSW solution with the difference that the day-night effect is
energy-independent.

The massless neutrino oscillations described in Eq. (6) also offer a
possible solution to solar neutrino observations.  There are two
solutions:  one at a small angle $(sin^2 2 \theta \sim 2.10^{-3}, \delta v/2 \sim
6.10^{-19})$ and one at a large angle $(sin^2 2 \theta \sim 0.7, \delta v/2
\sim 10^{-21})$\cite{mansour}.  This possibility can be tested in Long Baseline experiments.

The possibility of solar neutrinos decaying to explain the discrepancy
is a very old suggestion\cite{pakvasa}. 
The most recent analysis of the current solar neutrino data finds that
no good fit can be found:
$U_{ei} \approx 0.6$ and $\tau_\nu$ (E=10
MeV) $\sim 6$ to 27 sec. come closest\cite{acker}. The fits become acceptable if the suppression
of the solar neutrinos is energy independent as proposed by several
authors\cite{harrison} (which is possible if
the Homestake data are excluded from the fit).  The above conclusions are
valid for both decay scenarios A as well as B.

\section{Atmospheric Neutrinos}

The massless FCNC scenario has been recently considered for the
atmospheric  neutrinos by Gonzales-Garcia et al.\cite{gonzalez} with the matter effect 
supplied by the earth.
Good fits were found for the partially contained and multi-GeV events
with $\epsilon_q \sim 1, \epsilon'_q \sim 0.02$ as well $\epsilon_q \sim
0.08, \epsilon_q' \sim 0.07$.  The fit is poorer when higher energy
events corresponding to up-coming muons are included\cite{lipari}.  The expectations
for future LBL experiments are quite distinctive:  for MINOS, one expects
$P_{\mu\mu} \sim 0.1$ and $P_{\mu\tau} \sim 0.9.$ 
The story for massless neutrinos oscillating via violation of
Equivalence Principle or Lorentz Invariance is very similar.  Assuming
large $\nu_\mu - \nu_\tau$ mixing, and
$\delta v/2 \sim 2.10^{-22},$ a good fit to the contained
events can be obtained\cite{foot}; but as soon as multi-GeV and thrugoing muon
events are included, the fit is quite poor\cite{lipari}.

In both these cases the reason that inclusion of high energy upgoing
muon events makes the fits poorer is straightforward.  The up-going muon
events come from much higher energy $\nu_\mu's$ and the suppression is
weaker that in the lower energy events.  The energy dependence expected
in the above scenarios is very different:  no energy dependence in the
FCNC case and an averaging out of oscillations at high energy
(corresponding to 50\% uniform suppression) in the other scenario.

Turning to neutrino decay scenario A, it is found that it 
is possible to choose $\theta$ and $\alpha$ to provide a
good fit to the Super-Kamiokande L/E distributions of $\nu_\mu$ events and
$\nu_\mu/\nu_e$ event ratio\cite{discussion}.  
The best-fit values of the two parameters are $cos^2 \theta
\sim 0.87$   and $\alpha \sim 1 GeV/D_E,$ where
$D_E=12800$ km is the diameter of the Earth.  
This best-fit $\alpha$ value corresponds
to a rest-frame $\nu_2$ lifetime of
\begin{equation}
\tau_0 = m_2/\alpha \sim 
\frac{m_2}{(1 eV)} \times 10^{-10} s.
\end{equation}
Such a lifetime and decay length are not in conflict with current limits
on nonradiative modes.  Previous limits on $\nu_\mu$ decay lengths
from accelerators are only of the order of a few km.

It is possible to construct a viable decay model for Majorana neutrinos
with a massless Majoron which is predominantly singlet with an small
triplet admixture\cite{valle}.  The effective interaction will be given by
\begin{equation}
L=g \overline{\nu{_1^c{_L}}} \ \nu{_2{_{L}}} \ J \ + h.c.
\end{equation}

With this interaction, the rest-frame lifetime of
$\nu_2$ is given by
\begin{equation}
\tau_0 =
\frac{16 \pi}{g^2}
\frac{m_2^3}{\delta m^2 (m_1 + m_2)^2,}
\end{equation}
and hence, for the above fit, 
\begin{equation}
g^2 \delta m^2 \sim (2 -7) \times 10^{-4} eV^2
\end{equation}
Combining this with the bound on $g^2$ from $K, \pi$
decays\cite{barger1}, 
one finds
\begin{equation}
\delta m^2 \stackrel{>}{\sim} 0.73 \ eV^2.
\end{equation}
This result justifies the above approximation of large $\delta m^2$ and
the scenario A.  

We note in passing that with a choice of one $\delta m^2 \sim 
0(1) eV^2$ and another $\delta m^2 \sim 5.10^{-6} eV^2$ 
and a mixing matrix with large diagonal
entries and $U_{e2} \sim 0.02, U_{e3} \sim 0.04, U_{\mu 3} \sim 0.36$, it is
possible to account for all neutrino anomalies:  atmospheric, solar as well as LSND.

In this proposal, since the decays will be time dilated, at high
energies, the decay probability is very small and 
the suppression of $\nu_\mu$ flux is correspondingly smaller.  Indeed
it turns out that it is not possible to fit the observed angular
distribution of the high energy up-coming muon events\cite{lipari}.

If the neutrino that mixes with $\nu_2$ is different from the one that
it decays into, then the $\delta m^2$ in decay and the $\delta m^2$ in
oscillations are independent and it is possible for $\delta m^2$ in
oscillations to be very small.  In this case (scenario B), it is
possible to find an excellent fit to all the atmospheric neutrino data: 
contained events, multi-GeV events as well as upcoming muon
events\cite{barger}.  
The two decay possibilities lead to different expectations
for future Long Baseline experiments, e.g. for MINOS:  Case A leads to 
$P_{\mu \mu} \sim 0.8, P_{\mu \tau} \sim 0.2$ whereas case B leads to
$P_{\mu \mu} \sim 0.5, P_{\mu \tau} \sim 0.2$. 

Another proposal for explaining the atmospheric neutrinos is based on
decoherence of the $\nu_\mu's$ in the flux\cite{grossman1}.  The idea is that
$\nu_\mu's$ are interacting and getting tagged before
their arrival at the detector.  The cause is unknown but could be a
number of speculative possibilities such as a large neutrino background,
new flavor sensitive interactions in an extra dimension, violation of
quantum mechanics etc.  The $\nu_\mu$ survival probability goes as:
\begin{equation}
P= \frac{1}{2} \left [
1+ cos 2 \theta exp (-t/\tau) \right ]
\end{equation}
The Super-Kamiokande data can be fit by choosing $\tau \sim 10^{-2} s$
and $sin 2 \theta \sim 0.4$. A detailed fit to all the data over the
whole energy range has not been attempted yet.

\section{Conclusion}

As I mentioned at the beginning, the main motivation for this exercise
is to try to establish neutrino oscillations (due to mass-mixing) as the
unique explanation of the observed anomalies.  Even if neutrinos have
masses and do mix, the observed neutrino anomalies may not be due to
oscillations but due to other exotic new physics.  These possibilities
are testable and should be ruled out by experiments.  I have tried to show
that we are beginning to carry this program out.

\section{Acknowledgments}

I thank Milla Baldo-Ceolin for hosting this unique meeting and the
outstanding hospitality.  I thank Andy Acker, Vernon Barger, Anjan
Joshipura, Plamen Krastev, John Learned, Paolo Lipari, Eligio Lisi and
Tom Weiler for many enjoyable discussions and collaboration.

\end{document}